\documentclass[]{aa}
\usepackage{graphicx}
\usepackage{txfonts}
\usepackage{natbib}
\bibpunct{(}{)}{;}{a}{}{,}   

\begin{document}
   \title{The gas density around SN 1006}

   \author{F. Acero, J. Ballet, A. Decourchelle  }

   \institute{Laboratoire AIM, CEA/DSM-CNRS-Universit\'e Paris Diderot, DAPNIA/SAp, CEA-Saclay, 91191 Gif sur Yvette, France }

   \date{Received 2007 April 27; accepted 2007 August 17 }

  \abstract
   {The density of the ambient medium where the supernova remnant evolves is a relevant parameter for its hydrodynamical
       evolution, for the mechanism of particle acceleration, and for the emission at TeV energies.}
   {Using \textit{XMM-Newton} X-ray observations,  we present a study of  the ambient medium density of
    the historical supernova remnant SN 1006. }
   {We modelled the post-shock thermal emission to constrain the ambient medium density. Our study is focused on
   the North-West and the South-East rims of the remnant, where the thermal emission dominates.
   We used a plane-parallel shock
   plasma model plus another component for the ejecta that are not negligible in the regions of our study.
   The importance of the synchrotron component is also studied. In order to improve statistics, we
   combined several observations of the remnant.}
   {The density found in the South-East rim is low, roughly 0.05 cm$^{-3}$, and seems to be representative of the rest
   of the remnant. However, in the North-West rim (close to the bright optical filament), the density is significantly higher
   (about 0.15-0.25 cm$^{-3}$). This confirms a  picture of SN 1006 evolving in a
   tenuous ambient medium, except in the North-West where the remnant has recently encountered a denser region.}
   {A density this low is compatible with the non-detection of the remnant by the HESS gamma-ray observatory.
    The lower density in the South-East implies a higher shock speed of 4900 km/s, higher than that of 2890 km/s measured
    in the North-West.
 	This new estimate of the velocity could increase the maximum energy that accelerated particles can reach to
 	energies of about 1 PeV.}


		\maketitle

\section{Introduction}

The thousand-year-old Supernova Remnant (SNR) SN 1006 is a type Ia SNR located at a distance of 2.2 kpc
 \citep{wg03}.
The high Mach number collisionless shock of the remnant heats and ionizes the interstellar medium (ISM).
Furthermore, the blast wave is thought to be an efficient accelerator of cosmic-rays (CR).

Evidence that electrons are accelerated at the shock is found at several wavelengths.
The radio synchrotron  emission \citep{rg86} arises from electrons accelerated  to energies of at least 1 GeV.
 The X-ray synchrotron emission in SN 1006 \citep{ko95} comes from electrons of $\sim$ 20 TeV.
Evidence concerning proton acceleration is more scarce.
The best hope is to detect the  gamma-ray emission of pion decay resulting from the interaction of
accelerated protons (and heavier ions) with the thermal gas.
However, the TeV emission can also result from Compton scattering of  electrons at $\simeq$ 20 TeV on the CMB background.
 Currently there are only a few shell SNRs detected at those energies
  \citep[Cas A, Vela Junior, RX J1713.7-3946; ][]{ah01,ah05,ah06}  and SN 1006 is not yet one of them.

To interpret the TeV emission, it is important to estimate the gamma-ray flux from the  $\pi_{\mathrm{0}}$
mechanism. One of the most important parameters for predicting this flux is the gas density in the $\gamma$-ray emitting region.
 \cite{kb05} have shown that, in order to be consistent
with the non-detection of SN 1006 by the HESS experiment, the ambient medium density must be $\le$ 0.1 cm$^{-3}$.
The ambient medium density is  also a key parameter for the hydrodynamical evolution of the remnant.
As SN 1006 results from a type Ia SN, the surrounding medium is expected to be unaltered by the progenitor.
Moreover, the remnant is relatively isolated, as it is about 500 pc above the galactic plane. Thus it is
a  good  case study for comparing observations with theoretical models of CR  acceleration.

In the literature, the gas density around SN 1006 ranges from $0.05 \leq n_{\mathrm{H}} \leq 1.0 $ cm$^{-3}$.
The different estimates were based on radio H{\small{I}}   study \citep{dg02}, UV spectra \citep{lr96,kr04,rk07}, and
 X-ray emission \citep{hs86,wl97,lr03}. In SN 1006, while the limb-brightened synchrotron is dominant in the North-East
  and in the South-West,
 the thermal X-ray emission from the shocked gas dominates in the South-East (SE) and in the North-West (NW).
 It is possible to deduce the characteristics of this shocked gas, and in particular its density, with X-ray observations.
In the NW, such a study has already been carried out by \citet{lr03} with \textit{Chandra}.

Using \textit{XMM-Newton}, we present a study of the density in the South-East
and in the North-West by modeling the thermal emission.


\section{Data processing}

\begin{table}
 \caption{\textit{XMM-Newton} observations used in this paper}
 \label{tab:obs}
 \begin{tabular}{l|c|c c}

 \hline
 \hline
    &                 &      \multicolumn{2}{c}{MOS exp.(ks)}  \\
  Revolution\_ObsId & Observation Date &  Total & Good  \\

 \hline
 0305\_0111090601 (SE) &  2001 August 8 & 15.77 & 5.95 \\

  0306\_0077340101 (NW) &  2001 August 10 & 64.85 & 33.41  \\

  1044\_0306660101 (SE)&  2005 August 21 & 35.15 & 10.8  \\
 \hline
 \end{tabular}
 \end{table}

The first observations of the remnant's South-East region by the \textit{XMM-Newton} satellite \citep{rb04} were not very deep.
The effective exposure time in this region was 6 ks (ObsID 0111090601)
 in comparison with 33 ks for the North-West (ObsID 0077340101).
A second observation was carried out to provide better coverage of this region, on 2005 August 21 and 22 with
the medium filter for a duration of 35 ks.  Unfortunately, the proton flare contamination was high
and the remaining exposure time is only 11 ks. Note that for this new observation,
the CCD number 6 (no longer working) and the CCD number 4 (very noisy) of the camera MOS1 were not used.
 All observations used for our study  are listed in Table \ref{tab:obs} with their remaining
 exposure time after screening.
We used only MOS data due to its better spectral resolution. All the spectra used are the average of MOS1 and MOS2 detectors.

The data were processed using the Science Analysis System (SAS version 6.5).
All fitting  was carried out using unbinned spectra with the C-statistics implemented in Xspec (v12.2.1).
Binning was used for graphical purposes only and fixed at 3$\sigma$ for all spectra. We have fitted and
 plotted all the data from 0.3 keV to 10 keV.

 To clean proton flare contamination from the event files, we used the MOS-FILTER process available in the
 Extended Source Analysis Software (XMM-ESAS).
The following method was used for background subtraction. To take into account the astrophysical background,
 we extract a spectrum
 from a region outside of the SNR (regions A and B on Fig. \ref{fig:regions}); for statistical reasons, the area selected is as
 large as possible.
  The instrumental background spectrum is derived from a 960 ks blank sky observations compilation \citep{cr07}
in the same detector area as the source
and renormalized in the 10-12 keV band (where astrophysical photons are negligible) over the full field of view.

\section{Spectral analysis}
   \label{sect:spectral}
 To estimate the density of the ambient medium surrounding SN 1006, we model the thermal X-ray emission at the shock.
This gives us the post-shock density which is related to the gas density by
the compression ratio, which is 4 if particle acceleration is negligible.

\begin{figure}
   \centering
   \includegraphics[scale=0.45]{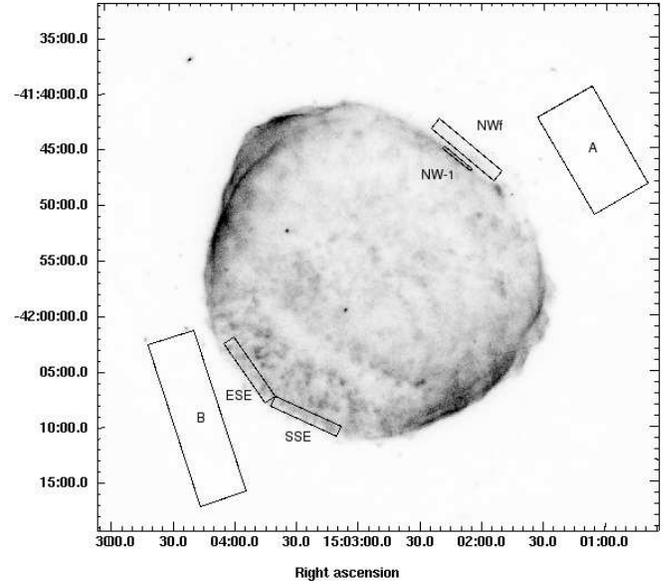}
   \caption{EPIC MOS plus PN image in the 0.5 to 2 keV. The scale is square root. The regions NW-1 and NWf are focused
   respectively on the bright and faint emission.
    The boxes named A and B represent the regions used for the astrophysical background for the NW and the
    SE respectively. }
   \label{fig:regions}
\end{figure}

Our study is focused on the SE and NW regions where thermal emission dominates.
Figure \ref{fig:regions} shows the extraction regions used for the spectra.
In the NW, we used two regions: one is the region noted NW-1 in the study of the remnant by \citet{lr03} with
 \textit{Chandra} (to relate our study to what has been done previously) and the second one is the faint emission region and is noted NWf.
Because the emission is faint in this region, we used a box as large as possible (1.2'x7.3'). The NW-1 region, in comparison, is
 much smaller (17"x3'). In the SE (region ESE and SSE), the width and length were fixed to 1' and 6.5' respectively, as a compromise between
 good statistics and being close to the shock.

\subsection{Method}
   \label{sect:method}

Our method is based on fitting the spectral data with thermal models and infering the post-shock density from the emission measure ($EM$)
 of the shocked ISM.
 Using Xspec, we get the parameter \textit{norm}=$10^{-14} \Omega /4\pi \int n_{\mathrm{e}} n_{\mathrm{H}} dl$
  which is proportional to the $EM$. In this expression, $\Omega$ is the
 solid angle of emitting gas contained in our extraction region and $dl$ the integration element along the line of sight.
 We assume that the shocked ISM fills all the volume inside the SNR. This implies a filling factor of 1.

  For a solar abundance plasma, the electron and hydrogen density are related by $n_{\mathrm{e}} = 1.21 \times n_{\mathrm{H}}$.
   Therefore, for a given value of \textit{norm} (obtained from the fit), we can solve
  the previous equation if we know the density profile to compute the integral.
We assumed a Sedov radial profile for that purpose.
   An important advantage of this method is that the hydrogen density is proportional to the square root
   of the ratio $norm/l$. Thus the density is weakly sensitive to variations of $EM$
    when using different thermal models, and to errors due to line of sight estimate.
    The error due to a filling factor of 1 is also minimized for the same reasons.

\subsection{What kind of model can we use ?}
   \label{sect:kindofmodel}

\begin{figure}
   \centering

  \includegraphics[bb= 0  0 514 764, clip,angle=-90,width=\columnwidth]{7742fig2.ps} \vspace{-3.9mm} \\
  \includegraphics[bb= 74 0 514 764, clip, angle=-90,width=\columnwidth]{7742fig3.ps} \vspace{-3.9mm} \\
  \includegraphics[bb= 74 0 590 764, clip, angle=-90,width=\columnwidth]{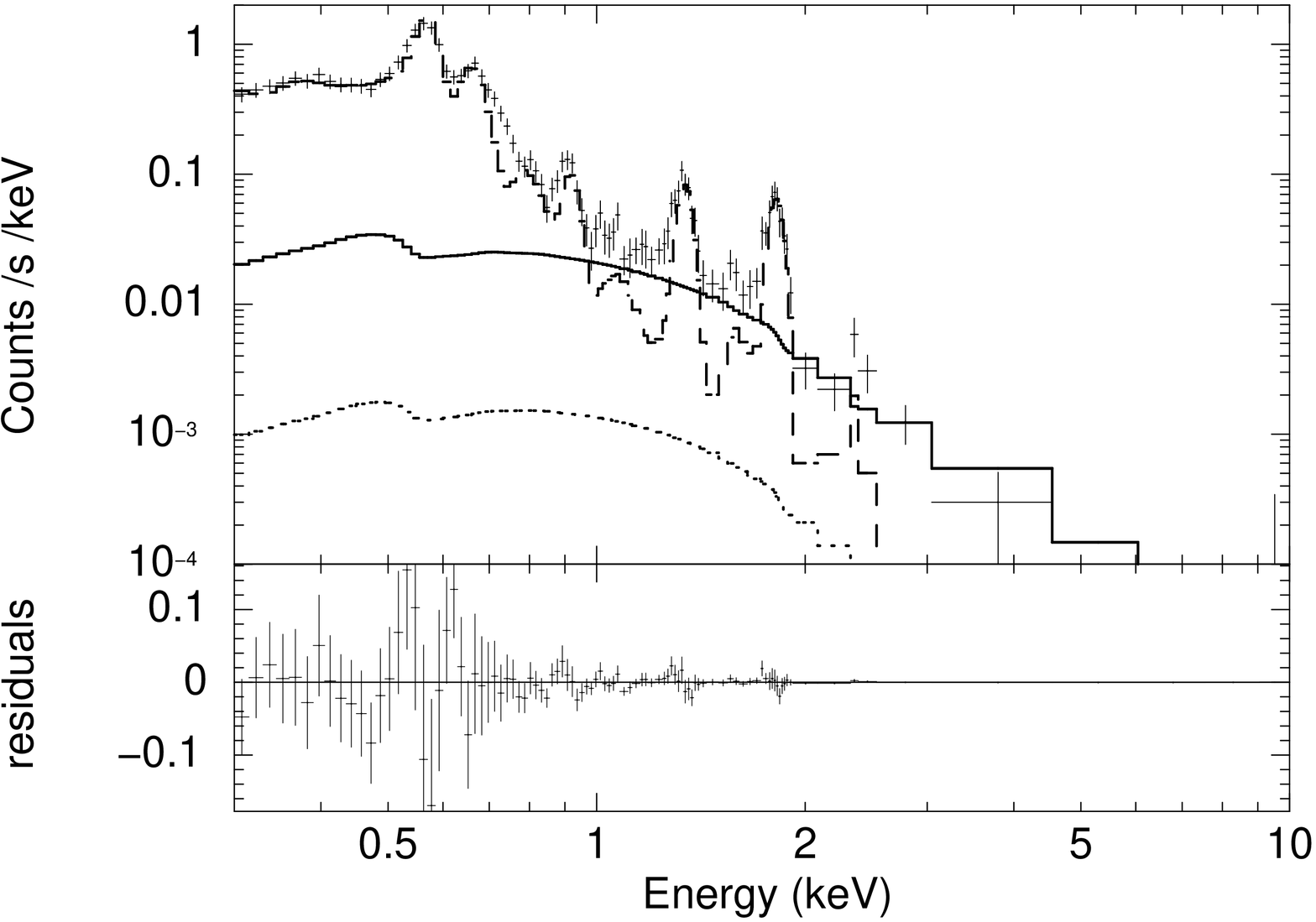} \vspace{-3.9mm}

   \caption{Best-fit spectrum from region ESE for the three models discussed.  \textit{Top}:
    shocked ISM. \textit{Middle}: shocked ISM (dotted line) plus ejecta model (dashed line).
     \textit{Bottom}: same model as above but with a synchrotron component (solid line).
    In the bottom panel, we note that the shape of the shocked ISM component is similar to the synchrotron component.
    The shocked ISM is almost not needed, and the power law shape is due to a very low ionization timescale $\tau \simeq 10^{8}$
    cm$^{-3}$s (no emission lines). For the middle and the bottom panel, the global fit (not shown) is a very good fit to the data.}
     \label{fig:3models}
  \end{figure}

The pertinence of our estimate of the density will depend on the quality of the spectral model for the shocked ISM.
As the ionization of the shocked ISM is out of equilibrium, we need to use non-equilibrium ionization (NEI) models.
Moreover, the variation  inside the SNR  of the ionization timescale, defined as $\tau  = \int n_{\mathrm{e}}dt$, is important.
 Starting at zero at the shock, the ionization timescale increases
 toward the center of the remnant down to $\sim \, 0.9 \, R_{\mathrm{shock}}$  for a Sedov model and then decreases.
Therefore we integrate cells of gas with different $\tau$ along the line of sight.

To model the emission, we used a simple model of plane-parallel shock
 (PSHOCK) with constant temperature, which allows the use of a
distribution of $\tau$ instead of a single ionization timescale like in a standard NEI model.
A plane-parallel shock model is a good local approximation for radially small regions close to the shock,
where $\tau$ is still increasing inwards.
The projection of the radial profiles onto the sky is not modelled by PSHOCK.

\begin{table*}
\begin{minipage}[t]{\columnwidth}
\centering
\caption{Best data fit parameters obtained for different models in region ESE.}
\label{tab:parameters}

\renewcommand{\footnoterule}{}  
\begin{tabular}{l | c c c c c c  }     
\hline

 &  & \multicolumn{3}{c}{Shocked ISM}  & Shocked ejecta & Synchrotron   \\
Parameters & C-stat & kT (keV) & $\tau$ (cm$^{-3}$s) & norm & kT & Cut-off frequency (Hz) \\

\hline

   One-component\footnote{A simple NEI model was also tried but gave a worse fit.} & 1266.1 & 1.77 & 3.8E9 & 4.9E-4 & ... & ... \\
   Two-component & 770.1 & 1.75 & 4.7E9 & 2.25E-4 & 0.52 & ... \\
   Three-component & 738.6 & 0.85 & 1.7E8 & 0.4E-4 & 0.62 & 4.5E16\\

\hline
\end{tabular}
\end{minipage}
\end{table*}

To check the consistency of our model, we also determine the density by comparing the maximum ionization timescale $\tau$ resulting
from the fit with the ionization timescale computed using Sedov solutions (refered to later as the $\tau$ method). Sedov solutions
have the advantage of taking into account the temporal evolution of the density and the shock velocity.
 At a given relative angular distance
$\theta / \theta_{\mathrm{shock}}$, the only parameter required is the age of the remnant, which is well known.
 The  method using the $norm$ parameter, which will be refered to later as the $EM$ method, is much more stable
 and will be used for the results. The $\tau$ method is useful for checking the consistency between the models we use.
In the following subsections, we use the region ESE as a test region to try different spectral models.

\subsection{Shocked ISM model}
   \label{sect:psh}

As a first step, we assumed that synchrotron emission is negligible and that all of the emission comes
 from thermal emission of the shocked ISM.
We therefore used a PSHOCK model with solar
abundances \citep{lo03} and we fixed the interstellar absorption to $\textit{N}_{\mathrm{H}} = 7 \times 10^{20}$ cm$^{-2}$ \citep{dg02}.
In the NEI model
(v1.1 in Xspec) the oxygen He-K $\delta$ and $\epsilon$ lines are not included
but they seem to be necessary to reproduce the large ``shoulder" of the He-like oxygen line complex.
To take them into account, we added one gaussian at 0.723 keV to represent the oxygen series  He-K $\delta$ and higher.

 The spectral fit of region ESE is shown in Fig. \ref{fig:3models} (top panel) and the resulting best-fit parameters are given
 in Table \ref{tab:parameters} (see ``one-component").
 The fit is poor and we clearly see that some emission lines like magnesium (at 1.35 keV) and silicon
  (at 1.86 keV)  cannot be reproduced by this model.
 Such emission lines require oversolar abundances and must come from the ejecta.

\subsection{Shocked ISM with ejecta model}
   \label{sect:pvpsh}

To reproduce the spectrum correctly, we added another component corresponding to the ejecta emission
 (the same PSHOCK model as above but with non-solar abundances).

To reduce the number of free parameters in this two-component model, we fixed most of the abundances of the ejecta.
As there is strong evidence that SN 1006 is a Type Ia SN,
we can obtain an estimate of the synthesized mass of elements and thus of the abundances in the ejecta through numerical
simulations \citep{ib99}.
 Since there is neither hydrogen nor helium in type Ia explosions, the abundances are given relative to oxygen
 with respect to cosmic abundances from \cite{lo03}.
We fixed the abundances of elements not constrained by emission lines in the spectrum to
the following values: O=1.0 (by definition); C=1.0; N=0.1; Ar=2.0; Ca=3.0.
Abundances from elements with prominent emission lines (Ne, Mg, Si) were let free.

 SN Ia are expected to produce an iron mass of 0.6 M$_{\odot}$, which is equivalent to a relative abundance of 20.
 This value is not compatible with our data because the iron emission lines are very faint in the observed spectrum.
  The value preferred by the fit is compatible with zero. In practice, we fixed it to Fe=0.1.
We can only see the emission from the shocked ejecta,
 and the spatial distribution of iron is complex \citep{wc01}. In our extraction region there appears to be little iron.

 The fit is much improved  in comparison with the one-component model with $\Delta$C-stat=495 (see Table \ref{tab:parameters}
 for C-stat and best-fit parameters).
 We note that the ejecta component is required for a good fit at Mg and Si emission
 lines, but that it also contributes to the emission at low energy and in particular to the oxygen lines.
In the first model, all the emission was attributed to the shocked ISM, whereas the emission of the ejecta is
 important, as seen in Fig.\ref{fig:3models} (middle panel).
Using the $EM$ method  for the shocked ISM component, we obtained a density $n_{\mathrm{H}}$=0.053 (0.048-0.057) cm$^{-3}$.
The value derived from the $\tau$ method is
consistent with the $EM$ method : $n_{\mathrm{H}}$=0.059 (0.05-0.07) cm$^{-3}$.
This model reproduces the spectrum  well and the two methods give a consistent estimate of the density.

\subsection{Impact of a potential synchrotron component}
   \label{sect:srcut}

SN 1006 is well known to be a non-thermal X-rays emitter. Most of the emission in the bright SW and NE limbs is
 due to synchrotron emission.
In the regions of our study, the thermal emission dominates at low energy, but can we completely neglect
the non-thermal emission?

In order to answer to that question, we built a three-component model by adding a SRCUT model.
We fixed the radio spectral index to 0.6 \citep{dr01}.
The radio flux was extracted from the same regions as the X-rays using the radio map from \cite{rb04}.
 This radio image is a combination of VLA data at the average frequency of 1517.5 MHz
and single-dish Parkes radiotelescope observations.

The spectral fit is improved with a $\Delta$C-stat=43.
The resulting cut-off frequency of $4.5\times 10^{16}$ Hz (about 200 eV) is compatible with
the frequency found by \citet{rb04} in their Fig. 7 (at a position angle 130 degrees).
 As we can see in Fig.\ref{fig:3models} (bottom panel), the synchrotron
component dominates the spectrum at high energy, at low energy, the shocked ejecta component dominates.

In the two-component model, the shocked ISM component was constrained by the spectrum at high energy, while the ejecta  were
constrained mainly by the Mg and Si emission lines.
Here, as the synchrotron reproduces the high energy, the shocked ISM is no longer well constrained.
In order to correctly reproduce the data, we need a continuum and a thermal model for the emission lines.
The continuum can be handled either by a high temperature thermal model or by the synchrotron model.
When we include both in the same fit, it is hard to disentangle them.
The emission lines can be more easily reproduced by the shocked ejecta as several abundances are left unconstrained.
Thus the shocked ISM component is almost not needed. The best-fit for a model with only synchrotron and shocked ejecta
has a C-stat that is higher by only $\Delta$C-stat=2.

Including the synchrotron in the model could lower the estimate of the density.
However, using hydrodynamical arguments we can set a lower limit on the density of the ISM, as explained later in Section
\ref{sect:rampressure}.
Therefore we think that the statistically preferred model is not the best physical model.

In conclusion, there are good physical reasons to include the synchrotron, but the data are not able to estimate its contribution.
We decided to use the simpler two-component model.
The resulting densities should be viewed as upper limits.

\subsection{ Using different sizes of extraction region }

 The width of our extraction regions ranges from 17" for region NW-1 up to 1' for the regions ESE and SSE.
We  checked the impact of varying the size of the extraction region on the density estimate.
In order to do so, we reduced the width of the region ESE
to 17", keeping it immediately behind the shock.
 Using the two-component model described in Sect. \ref{sect:pvpsh}, we derive densities  from the
small ESE region with the $EM$ and $\tau$ methods of 0.065 (0.055-0.075) cm$^{-3}$ and 0.06 (0.048-0.09) cm$^{-3}$ respectively.
The density estimate is fully consistent with that determined with a larger ESE region (see Sect \ref{sect:pvpsh}).
 Therefore the density estimate is not sensitive to the size of the extraction region.

\section{X-ray results}
   \label{sect:results}

\begin{table*}
 \caption{Best-fit parameters obtained with the shocked ISM plus shocked ejecta model (two-component) for the different regions.
  The gas density derived with the $EM$ and the $\tau$ methods is also given (see Sect. \ref{sect:method}).
   Due to the faint emission in region NWf,
  the temperature and the abundances of the shocked ejecta component (parameters in square brackets)
  were fixed to the values of region NW-1. The lower and upper bounds are given in parentheses at a 90 \% confidence level.
  * : The density estimate for region NW-1 assumes constant density and velocity rather than Sedov profiles
  (see Sect. \ref{sect:NW-1}).}
 \label{tab:results}
 \centering
 \begin{tabular}{l|c c c c}

 \hline
 \hline
  Region                          &  NW-1    &   NWf    &   SSE   &   ESE  \\

 \hline

 k$T_{\mathrm{ejecta}}$ (keV)                 &  0.12 (0.10-0.13)    &   [0.12]     &   0.55 (0.5-0.65)   &   0.52 (0.49-0.57)  \\

  Ne/O                 &  14  (8-22)  &   [14]    &   0.24 (0.1-0.3)  &   0.22 (0.1-0.3)  \\

  Mg/O                &  49 (25-80)    &   [49]    &   6.0 (4.0-8.0)   &   4.7 (4.0-5.5)   \\

  Si/O                &  21 (5-40)    &   [21]    &   19.0 (17.0-23.0)   &   17.0 (14.0-20.0)   \\

  \hline

  k$T_{\mathrm{ISM}}$ (keV)                    &  1.45 (1.55-2.1)     &   1.75 (1.5-2.1)    &   2.1 (1.6-2.5)   &  1.75 (1.41-2.15)  \\

 $\tau$ of shocked ISM ($10^{9}$ cm$^{-3}$ s)    &  4.45 (3.5-5.6)   &   2.4 (2.15-2.7)    &   6.0 (5.4-6.5) &   4.7 (4.0-5.6) \\

$norm$ ($10^{-4}$ cm$^{-5}$)         &  0.55 (0.4-0.75)    &   2.56 (2.2-2.9)    &   1.88 (1.95-2.5)   &   2.25 (1.8-2.6)   \\

    \hline
  C-statistics         &    &    &   &     \\
  Two-component model                             &  454.8    &   814.3    &   750.1   &   770.1   \\
  One-component model                         &  646.6   &  840.5    &   1210.4   &   1266.1   \\

   \hline
  Gas density ($10^{-2}$ cm$^{-3}$)   &    &    &   &     \\
  $\tau$ method &  15$^{*}$ (11.2-18.5)  &  3.0 (2.6-3.3)  & 7.5 (6.8-8.1)  &  5.9 (5.0-7.0)   \\
  $EM$ method  &   -  &  4.9 (4.5-5.2)  & 5.2 (4.9-5.6)  &  5.3 (4.8-5.7)   \\

 \hline
 \end{tabular}
 \end{table*}

\begin{figure*}
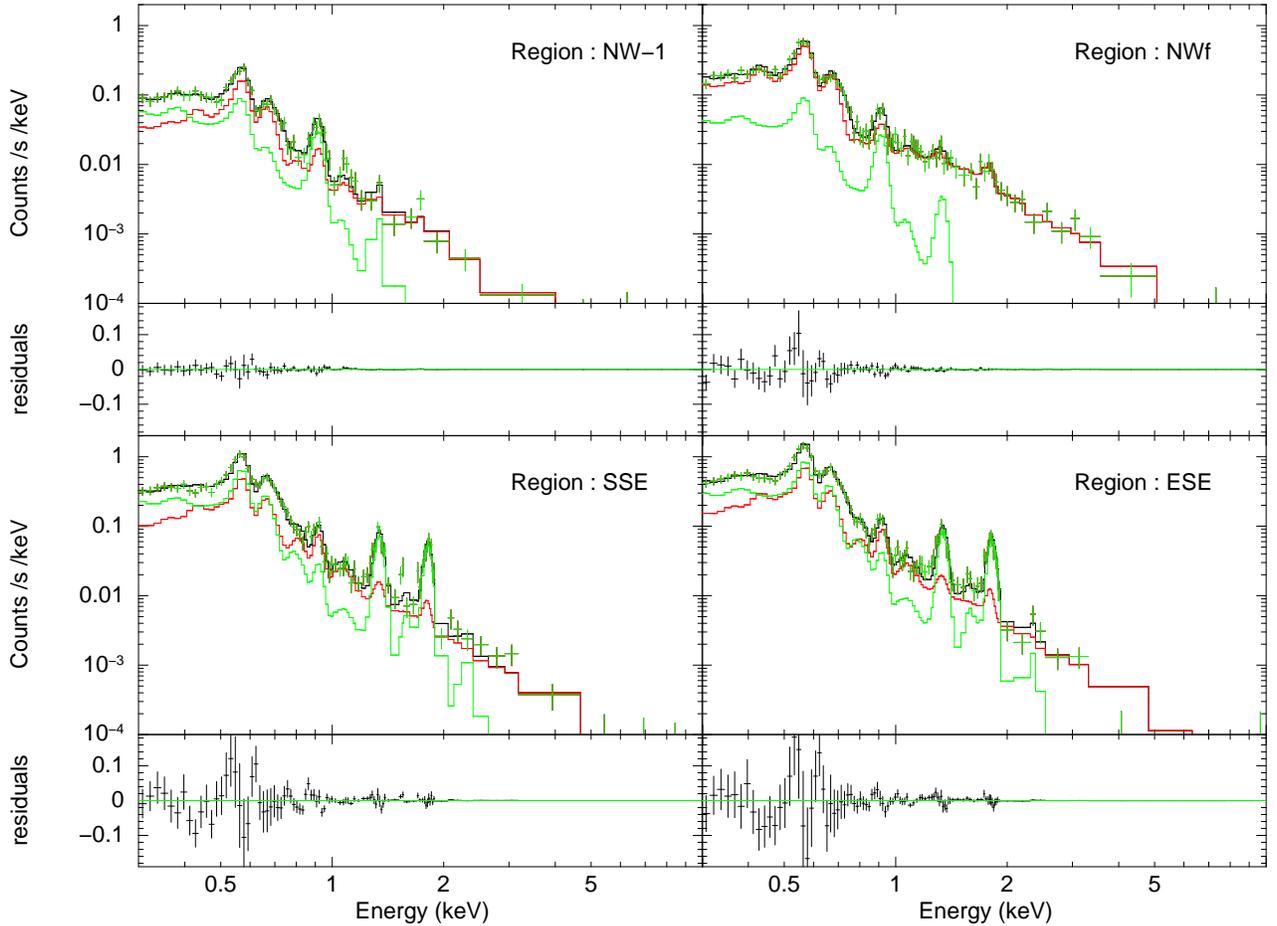

   \centering
   \begin{tabular}{cc}

 { \includegraphics[bb= 74  -10 514 704, clip,angle=-90,scale=0.37]{7742fig5.ps} } & \hspace{-7mm}

 { \includegraphics[bb= 74  127 514 704, clip,angle=-90,scale=0.37]{7742fig6.ps} } \vspace{-12.2mm} \\

 { \includegraphics[bb= 0  -10 575 704, clip,angle=-90,scale=0.37]{7742fig7.ps} } & \hspace{-7.8mm}
 { \includegraphics[bb= 0  127 575 704, clip,angle=-90,scale=0.37]{7742fig8.ps} }

   \end{tabular}

   \caption{ Best spectral fit  for the four regions with the double-component model (see Sect. \ref{sect:pvpsh}).
    The global fit is shown in black, the shocked ISM component in red (dominant at high energy), and the shocked ejecta in green. The 0.723 keV line
    (Sect. \ref{sect:psh}) is included in the global fit. }

   \label{fig:4spectrum}
\end{figure*}

\subsection{General results}

Following the spectral analysis of region ESE (Sect. \ref{sect:spectral}), we now use the same model for all the regions:
 shocked ISM plus shocked ejecta (as in Sect. \ref{sect:pvpsh}).
 The results for the four regions (ESE, SSE, NWf, NW-1) studied here are listed in Table \ref{tab:results}.

 For all the regions, the ejecta component has greatly improved the quality of the spectral fit in comparison with a one-component
 model. The emission lines are then well reproduced with our model as shown in Fig. \ref{fig:4spectrum}.

The density obtained with the $EM$ method in the SE rim (ESE and SSE region) and in the  NW faint region (NWf)
  is globally low  ($n_{\mathrm{H}}$=0.049-0.053 cm$^{-3}$). Moreover the plasma parameters of the shocked ISM obtained from NWf
are similar to those found in the SE.
This is consistent
 with the smooth appearance of the H$\alpha$ image \citep{wg03}.
We also note that the temperature of the shocked ISM
is comparable in those three regions: k$T_{\mathrm{ISM}}$=1.75-2.1 keV.
This ISM temperature is much higher than the temperature
of the shocked ejecta (0.5 keV). The shocked ISM component dominates the spectrum at high energy.

As we can see on the spectra in  Fig. \ref{fig:4spectrum},
 the ejecta component is important below 2 keV.
 Its contribution to the Mg, Si and Ne emission lines is obvious, and we can  also see that part
 of the emission from the oxygen lines is produced by shocked ejecta.
In the NW regions, an overabundance of Ne
is required. This has also been noted by \citet{lr03} in their spectral analysis of the bright filament.

\subsection{The bright filament}
\label{sect:NW-1}

The bright filament seen in X-rays in the NW is coincident with a bright H$\alpha$ filament.
A possible explanation is a ``recent" interaction of the expanding SNR with a denser ISM.
The sudden lower shock velocity flattens the shape of the blast wave in this region as it is seen in optical and X-rays.
The thermal emission at the shock increases because the shock propagates a denser region and thus creates a
brighter filament than in the SE.
The fact that the shock is entering in a denser region in the NW is particularly visible in H$\alpha$.
The radio study by \cite{dg02} also detected a higher density in the NW.
In H$\alpha$ and in radio, we also see the faint structure of the region NWf.
A possible interpretation of this faint border is that the remnant is expanding
in front of (or behind) the bright filament in a region of lower density similar to that of the SE.

 To derive the density from the plasma parameters, we used a simple Sedov model to compute
 $\tau$ and $EM$ and then compare them with the best-fit parameters (see Sect. \ref{sect:method}).
However, in the NW-1 region it seems that
part of the shock wave has recently encountered a denser region. For this reason, the geometry of the shell is too complex
 and probably not well represented by a Sedov density profile. The density derived with this method is probably
not reliable in this particular region.
 Nevertheless, with a simpler model (assuming constant density and constant velocity)
 we can derive  the density from the ionization timescale \citep{lr03}.
  Indeed, we know the shock speed in this region \citep[2890 km/s,][]{wg03} and  know that region NW-1  has a radial extent
of 17" (corresponding to an elapsed time of 240 years since the gas was shocked, at a distance of 2.2 kpc).
Thus for $\tau$=4.5 (3.5-5.5) $\times 10^{9}$ cm$^{-3}$ s,
the density is $n_{\mathrm{H}} \simeq 0.15 \, (0.11-0.18)$ cm$^{-3}$. We note that k$T_{\mathrm{ISM}}$ is lower in region NW-1 (1.45 keV)
than in the SE and NWf (1.7-2.1 keV). This is consistent with the fact that the density in region NW-1 is higher.

In comparison, the study of the bright filament region by \citet{lr03}  with \textit{Chandra} found a density of  0.25 cm$^{-3}$.
 To see if there are differences between the data observed by \textit{XMM-Newton} and \textit{Chandra},
 we tried the same simple solar-abundance PSHOCK model  \citep[described in Sect. 3.2 of][]{lr03} on our data.
 The absorption column density they use is higher than our value:
8.9 vs 7.0 $ \times 10^{20}$ cm$^{-2}$. We modelled the data with both values.
 The best-fit parameters  are given in Table \ref{tab:chandra}. The model with low $\textit{N}_{\mathrm{H}}$
 is statistically preferred with a $\Delta$C-stat of 35.

\begin{table}[!ht]
 \caption{Comparison of the best-fit parameters obtained with \textit{XMM-Newton} and \textit{Chandra} data in the NW-1 region.}
 \label{tab:chandra}
  \centering
 \begin{tabular}{l|c c c c}
\hline
 \hline
           &      \multicolumn{2}{c}{k$T_{\mathrm{e}}$ (keV) }    &      \multicolumn{2}{c}{$ \tau \, (10^{9} \mathrm{cm}^{-3}$ s) }   \\
	& XMM & Chandra &  XMM  & Chandra                                 \\
  $\textit{N}_{\mathrm{H}}$=8.9 $ \times 10^{20}$ cm$^{-2}$  & 0.6 & 0.6 & 6.8 & 6.9 \\
  $\textit{N}_{\mathrm{H}}$=7.0 $ \times 10^{20}$ cm$^{-2}$  & 0.7 & ... & 6.5 & ... \\

 \hline
 \end{tabular}
 \end{table}

Both data give similar parameters and the fact that we are not using the same value for $\textit{N}_{\mathrm{H}}$
does not have an important impact on the plasma parameters. Thus, the  difference in the estimate of the density
is not due to the data;  it is mostly due to the difference of models.
When taking into account the ejecta using a two-component model, the density derived is slightly lower than
 when using a simple solar abundance PSHOCK model.

The difference of density between the NW-1 and other regions was already pointed out by \cite{kr04} and \cite{lr03}.
 \cite{kr04}  estimated the density in the NE to be  $\leq$ 0.06 cm$^{-3}$, which is compatible with what we found in the SE.

We conclude that a low value of $\sim$ 0.05 cm$^{-3}$  is probably representative  of the ambient medium around SN 1006,
except in the bright NW filament where the density is higher  (0.15-0.25 cm$^{-3}$, depending on the model).

\section{Discussion}

\subsection{Comparison with other observations}
\label{sect:otherobs}

By modelling the post-shock thermal emission of the remnant in the SE and in the NW, we found a density of $\sim$ 0.05 cm$^{-3}$.
Many other studies (local and global) have been carried out to measure this  density and provide estimates
  in the range of 0.05 cm$^{-3}$  to 1.0  cm$^{-3}$.

Most of the local ones were focused on the bright Balmer line filament in the NW. Using HST observations of this filament,
  \citet{rk07} derived a density
  in the range of 0.25-0.4 cm$^{-3}$ by modelling the H$\alpha$ brightness behind the shock.
By using the high spatial resolution X-ray observatories, some studies also estimated the density in the NW region.
\cite{wl97}, with \textit{ROSAT}, used a model for the X-ray surface brightness profile behind the shock;
they found a density of 1.0 cm $^{-3}$.
However, using the \textit{Chandra X-ray Observatory}, \citet{lr03} studied the same region and found a much lower value.
They derived a density of 0.25 cm$^{-3}$ from the ionization timescale in a region right behind the shock.
The spectral analysis of the same region using \textit{XMM-Newton} observations gave an identical value
when using the same model (see Sect. \ref{sect:NW-1}). A model taking into account the ejecta component gave a slightly lower
value (0.15 cm$^{-3}$). We note that the density derived from the X-rays is lower than that derived from the optical.
By using X-ray observations, we average the density on the size of the extraction
region, whereas in optical we measure the density right at the shock. If the shock wave is entering into a cloud,
it is plausible that the difference of density between optical and X-rays is related to the gradient of density
when entering the cloud.

There is  a general agreement that the remnant is encountering a denser region in the NW,
so the density there is not directly comparable with what we found in the SE.

The global studies  average the density all around the remnant.
A surrounding density of $\sim$ 0.3 cm$^{-3}$ has been inferred by \citet{dg02} with a survey of the H{\small{I}} emission.
However, this result is the average density
along a line of sight of several hundreds of pc, which is very large in comparison with the typical size of the remnant
of 20 pc.

 \citet{hs86} derived a density of  0.05  cm$^{-3}$
  by modelling the global X-ray spectrum under the assumption
 that most of the emission is thermal.
Spatially resolved spectral analysis showed that this assumption is only valid locally in the NW and in the SE.
The global spectrum is in fact dominated by synchrotron emission.
Therefore, even though their result agrees with ours, we view this as coincidental.
Numerical simulations have also been carried out to study the evolution of SNRs in constant density ISM.
In order to be consistent with the observed angular size and expansion rate of SN 1006, \citet{dc98} derived from
their simulation a density between 0.05 cm$^{-3}$ and 0.1  cm$^{-3}$. Those results are compatible with what we found.
We note that our density is lower than values of most other studies, but most of them were focused on the NW,
where the ambient medium seems denser.
The value found in the SE region is more likely to be representative of the overall ambient medium density, including
in the bright synchrotron limbs.

\subsection{Limitations of our model}

 Our method is based on estimating the density of  the shocked ambient medium. To infer the density of the unshocked
 ambient medium, we have to assume a value for the compression ratio.
 In the efficient acceleration sites (NE and SW rims) the compression ratio can increase to 6 as
 calculated in \citet{bk02}. However, in the SE and NW rims, the acceleration process is not as efficient, as we see
 very little X-ray or radio
 synchrotron emission. Thus a compression ratio of 4 seems reasonable.

To derive the density from the $EM$, we assumed that the shocked ISM filled all the volume i.e. a filling factor
$\varepsilon$=1. As  seen in Sect. \ref{sect:method}, we need ejecta to reproduce the spectra and it
  must occupy some fraction of the volume.
 In the ejecta-dominated phase \citep{ch82}, the fraction of the volume occupied by the shocked ISM in
the global shocked region
is about 0.8. SN 1006 is in a more advanced phase (towards Sedov) where the ejecta are more diluted in the shocked ISM.
Moreover, closer to the shock,
as in our regions, $\varepsilon$ is supposed to be closer to unity.
So in our model, we can reasonably say that $\varepsilon = 0.8$ is a lower limit to the filling factor.
In that case, the density estimate for region ESE would increase from
0.053 cm$^{-3}$ to 0.059 cm$^{-3}$.

To derive the density from the $norm$ parameter, we assumed a Sedov density profile to estimate the value of the
 integral $\int n_{\mathrm{e}}^{2}dl$.  As SN 1006 is switching from the ejecta-dominated
stage to the Sedov phase, the density profile that we assumed is steeper than it is in reality.
We therefore overestimate the density corresponding to a given value of the $EM$.
In the extreme case of a flat density profile, we obtain for the ESE region a density of 0.042 cm$^{-3}$ in comparison
with 0.053 cm$^{-3}$ for the Sedov profile.

The distance of 2.2 kpc used throughout this paper was derived by \citet{wg03} from the  measurement of the shock speed
of the bright filament of 2890$\pm$100 km/s and the apparent expansion of 0".28 yr$^{-1}$.
 However, \citet{hm07}, using the same
data as in \citet{wg03} but with a new model that incorporates additional physics, derived a lower
shock speed of 2509$\pm$111 km/s. The corresponding distance to the remnant with this shock speed is 1.9 kpc.
For our study this will change the length of the line of sight in the $EM$ method, but it will have a low impact
on the density as $n_{\mathrm{H}} \propto \sqrt{norm/dl}$.

Concerning the contribution of the synchrotron emission, our three-component model was not able to disentangle
the thermal emission of the shocked ISM from the synchrotron emission.
We cannot exclude that part of the emission at high energy is due to synchrotron emission, so our
density estimate is an upper limit in that respect.
A lower limit to the density is discussed later in Sect. \ref{sect:rampressure}.

\subsection{Velocity of the blast wave}
\label{sect:properties}

With the density of 0.05 cm$^{-3}$ inferred above, we can reestimate several parameters in order to account for
the observed  radius of 9.6 pc (at a distance of 2.2 kpc). The age is t=1000 years, and as
SN 1006 seems to be a type Ia supernova, the ejecta mass is M$_{\mathrm{ej}} = 1.4$ M$_{\odot}$.
The impact of the distance uncertainty was low in the density estimate. However,  parameters like energy or shock
speed are really dependent on this distance. The following results are given for both d=2.2 kpc and 1.9 kpc.
As the remnant is just switching from the ejecta-dominated stage to the Sedov stage, we decided to use
the equation from \cite{tm99} for an homogeneous ISM with s=0 and a type Ia ejecta profile with n=7.

The kinetic SN energy consistent with the previous parameters is E$_{\mathrm{SN}}$=2.2/1.3$\times 10^{51}$ ergs.
For example, for a density higher than 0.2 cm$^{-3}$, the energy required would be higher than
 E$_{\mathrm{SN}}$=6.0/3.3 $\times10^{51}$ ergs.
  Type Ia SN are expected to release an amount of energy of $\sim$ 10$^{51}$ ergs.
  Energies above 2$\times$10$^{51}$ ergs seem  hard to reach.
  Thus a uniform gas density $\geq$ 0.2 cm$^{-3}$ surrounding all the remnant is excluded for energetic reasons.

The global velocity of the blast wave can also be reestimated.
As the density is lower in the SE, the shock speed there must be higher than the shock speed in the NW which is 2900 km/s.
Indeed, the shock speed derived from the Truelove \& McKee equations  with
 an ambient medium density of 0.05 cm$^{-3}$  at t=1000 years is  4900/4600 km/s.
If the density found in the SE region is representative of the rest of the remnant, this velocity is what we can expect
for the whole remnant except in the NW. We note that there is also a difference between the global shock speed
measured in radio and the velocity in the NW deduced from H$\alpha$ observations.
 Indeed, even though the error bars are big, the mean proper motion measured by \citet{mg93}
 in radio is 0".44$\pm$0".13 yr$^{-1}$, corresponding to 4500 km/s for a distance of 2.2 kpc.
 This is much higher than the velocity of 2890 km/s obtained in the NW, and it is compatible with our estimate.

\begin{figure}
   \centering

 \includegraphics[width=\columnwidth]{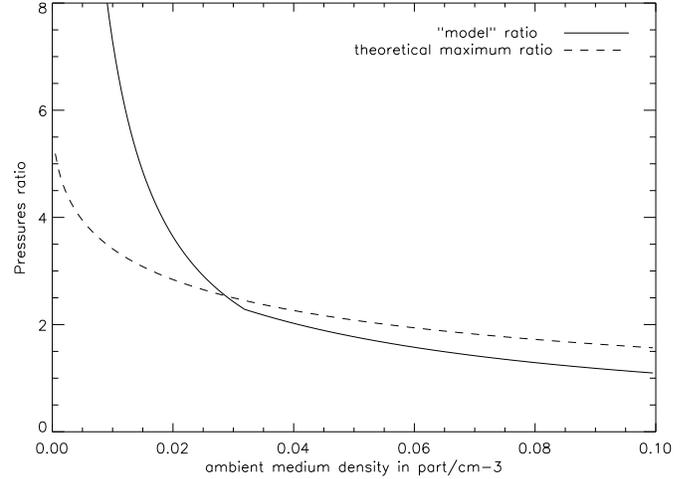}
   \caption{Theoretical and ``model" ratio of  pressures behind the transmitted shock in the NW cloud and behind the blast wave
     elsewhere.}
     \label{fig:ratio}
  \end{figure}

\subsection{Lower limit on the density}
\label{sect:rampressure}

When a shock wave encounters a cloud, a transmitted shock is driven into the cloud.
The  pressure behind the transmitted shock is greater, at most, by a factor $\beta$ than the  pressure
behind the blast wave not in interaction with the cloud.
For a given density contrast n$_\mathrm{cloud}$/n$_\mathrm{ISM}$, we can derive this $\beta$ factor using the equations
from \citet{sg75}.
The  pressure is proportional to $n_{\mathrm{H}}$.$V_{\mathrm{shock}}^{2}$.
What we are interested in here are the density and velocity right at the shock.
We know those values in the NW (transmitted shock) from the optical studies : $n_{\mathrm{H}}=0.25-0.4$ cm$^{-3}$
 and $V_{\mathrm{shock}}^{2}$=2890 km/s \citep[][respectively]{rk07,wg03}.
For any ambient medium density,
the velocity (in a region where the shock is not encountering any cloud) is known using the Truelove \& McKee equations.
 Therefore we can derive a ``model" ratio of  pressures for any ambient medium density.
By comparing the theoretical ratio $\beta$ with the ``model" ratio, we can set a lower limit to
the ambient medium density. The change of slope in the ``model" ratio corresponds to the transition between the ejecta-dominated
phase and the Sedov phase.
For n$_\mathrm{cloud}$=0.25 cm$^{-3}$, the ``model" ratio becomes higher than the maximum
theoretical value of $\beta (0.25/n_\mathrm{ISM})$ for n$_\mathrm{ISM}$ lower than 0.03 cm$^{-3}$ (see Fig. \ref{fig:ratio}).
For n$_\mathrm{cloud}$=0.4 cm$^{-3}$, the
lower limit on the ambient medium density is 0.07 cm$^{-3}$.
We conclude that the density of the ambient medium where the SNR evolves cannot be lower than 0.03 cm$^{-3}$.
For our preferred value of the density of 0.05 cm$^{-3}$, we have a ``model" ratio of 1.7 which is compatible with
the theoretical value of $\beta(0.25/0.05)$=2.1.

\subsection{Implication for CR acceleration}

The efficiency of the acceleration of the particles is strongly dependent on the shock speed.
In SN 1006, this velocity has been measured only in the NW region, by \cite{gw02} and \citet{hm07}.
However, the velocity there does not
seem to be representative of the rest of the remnant, and in particular, of the efficient acceleration sites
 in the NE and SW rims. If we use for those regions the velocity estimated in the SE in Sect. \ref{sect:properties},
this has an impact on the highest energy the accelerator can reach.
In fact, in a Bohm regime, at an age of 1000 yr we have
$E_{\mathrm{max}} = 32.7 \frac{B_{\mathrm{d}}}{100 \, \mathrm{\mu G} } (\frac{V_{\mathrm{shock} }}{1000\,\mathrm{km/s}})^{2}$
TeV  \citep[adapted from][]{pm06} where $B_{\mathrm{d}}$ is the downstream magnetic field. Thus with the higher velocity
 (4900 km/s) and $B_{\mathrm{d}}$=150 $\mu$G as in \cite{kb05},
  the maximum energy that particles could reach in the bright limbs would be $E_{\mathrm{max}}\simeq$ 1 PeV.

The TeV emission of those accelerated particles depends on the density of the ambient medium.
The total $\gamma$-ray flux (IC+$\pi_{\mathrm{0}}$-decay) from the NE half of the remnant
 for different values of $n_{\mathrm{H}}$ has been given in \cite{kb05}. For an upstream magnetic field of 30 $\mu$G,
  a density of 0.05 cm$^{-3}$ and a distance of d=2.2 kpc, this flux is a factor of 3 below the current HESS upper limit, and
  only a factor $\sim$ 2 for d=1.9 kpc.
This would, however, need to be revisited in the light of our revised estimate of the shock velocity.

\subsection{Electron temperature equilibration}
A  question that remains open in the physics of strong shocks is the mechanism by which the electrons are heated just behind
the shock. In the Rankine-Hugoniot relations
for  high Mach number shocks, the particles are heated in relation to their mass. The proton temperature is higher
 than the electron temperature by a factor 1836. However, it has also been proposed by \cite{cp88} and \cite{gl07}
that collective plasma interaction can give rise to prompt (limited) electron heating at the shock.
Such mechanisms produce a higher electron temperature at the shock than in the Rankine-Hugoniot case.
Our model requires an electron temperature k$T_{\mathrm{e}} \sim 2$ keV to reproduce the spectrum in the region ESE.
 The Rankine-Hugoniot relations
(without particle acceleration) imply a proton temperature just at the shock of $\sim$ 40 keV
for a shock speed of 4900 km/s,
if the degree of temperature equilibration is low. Using the Coulomb equilibration mechanism
with an initial electron-to-proton temperature ratio of 1/1836 (the electron-to-proton mass ratio),
the predicted electronic temperature is  lower than what is observed.
Indeed, with an ionization timescale of $5\times 10^{9}$ cm$^{-3}$ s, k$T_{\mathrm{e}}$ rises to 0.8 keV only.
An initial ratio of at least $T_{\mathrm{e}}/T_{\mathrm{i}}$=0.04 (corresponding to k$T_{\mathrm{e}}$=1.7 keV just at the shock) is needed for
the temperature to reach 2 keV.
According to  \citet{gl07}, the electronic temperature immediately behind
the shock is constant ($\sim$ 0.3 keV) for any shock speed above 400 km/s.
In comparison, the electronic temperature just at the shock required in our model  (k$T_{\mathrm{e}}$=1.7 keV) is much higher.

However, we note that this high temperature depends very much on the amount of synchrotron emission
in the SE, as discussed in Sect. \ref{sect:srcut}. To check if the three-component (synchrotron, shocked ejecta and shocked ISM)
model is compatible with a low electronic temperature just at the shock, we fixed several parameters of
the shocked ISM component for the lowest possible value of the ambient medium density  i.e. 0.03 cm$^{-3}$.
The $norm$ parameter and the upper limit of the ionization timescale can be fixed from Sedov models and from the geometry
of the emitting region. They are fixed respectively to 0.8$\times 10^{-5}$ cm$^{-5}$ and
$3\times 10^{9}$ cm$^{-3}$ s. From the upper limit of $\tau$, we can derive  an upper limit to the temperature
but not the average (for the region) electron temperature that is needed for the PSHOCK model.
In order to avoid this difficulty, we use a plane-parallel shock model with separate ion and electron temperatures (NPSHOCK)
 to reproduce the shocked ISM. In this model the two temperatures can be set precisely.
 The mean post-shock temperature is derived from the shock speed (22 keV at 4900 km/s),
  and the electron temperature immediately behind the shock front is set to 0.3 keV.
The global fit is good and the C-stat is higher only by $\Delta$C-stat=5 in  comparison to the three-component model
 (where the parameters for the shocked ISM component were left free). Thus the data are
compatible with a low electron temperature just behind the shock if a model including synchrotron emission is assumed.

\section{Summary}

The study of the post-shock thermal emission in the SE and in the NW of SN 1006 with \textit{XMM-Newton}
observations leads to the following conclusions :

\begin{enumerate}

\item  Most previous studies were focused on the NW region where the remnant seems to be encountering a denser ISM.
By modelling the post-shock thermal X-ray emission, we were able to estimate the density in the SE rim.
 The low value of $\sim$ 0.05 cm$^{-3}$ we found seems representative of the rest of the remnant.

\item    The shock speed in the SE region, and maybe in the rest of the remnant, is $\sim$ 4900 km/s. The velocity of
2890 km/s was deduced from UV observations in the NW where the density is higher, causing the expansion to slow down.

\item The highest energy particles can reach scales as $V_{\mathrm{shock}}^{2}$. The new estimate of the velocity could
 increase E$_{\mathrm{max}}$ up to energies of about 1 PeV.

\item If we assume that the ambient medium density found in the SE region extends to the bright limbs where the acceleration
is efficient, then according to \citet{kb05} the TeV gamma-ray flux would be a factor of $\sim$ 3 below the current HESS upper limit.

\end{enumerate}

\begin{acknowledgements}
We are grateful to the referee for his valuable suggestions which helped us to improve the clarity of the paper.
\end{acknowledgements}


\begin{thebibliography}{31}
\expandafter\ifx\csname natexlab\endcsname\relax\def\natexlab#1{#1}\fi

\bibitem[{{Aharonian} {et~al.}(2001){Aharonian}, {Akhperjanian}, {Barrio},
  {Bernl{\"o}hr}, {B{\"o}rst}, {Bojahr}, {Bolz}, {Contreras}, {Cortina},
  {Denninghoff}, {Fonseca}, {Gonzalez}, {G{\"o}tting}, {Heinzelmann},
  {Hermann}, {Heusler}, {Hofmann}, {Horns}, {Ibarra}, {Iserlohe}, {Jung},
  {Kankanyan}, {Kestel}, {Kettler}, {Kohnle}, {Konopelko}, {Kornmeyer},
  {Kranich}, {Krawczynski}, {Lampeitl}, {Lopez}, {Lorenz}, {Lucarelli},
  {Magnussen}, {Mang}, {Meyer}, {Mirzoyan}, {Moralejo}, {Ona}, {Padilla},
  {Panter}, {Plaga}, {Plyasheshnikov}, {Prahl}, {P{\"u}hlhofer}, {Rauterberg},
  {R{\"o}hring}, {Rhode}, {Rowell}, {Sahakian}, {Samorski}, {Schilling},
  {Schr{\"o}der}, {Siems}, {Stamm}, {Tluczykont}, {V{\"o}lk}, {Wiedner}, \&
  {Wittek}}]{ah01}
{Aharonian}, F., {Akhperjanian}, A., {Barrio}, J., {et~al.} 2001, \aap, 370,
  112

\bibitem[{{Aharonian} {et~al.}(2005){Aharonian}, {Akhperjanian}, {Bazer-Bachi},
  {Beilicke}, {Benbow}, {Berge}, {Bernl{\"o}hr}, {Boisson}, {Bolz}, {Borrel},
  {Braun}, {Breitling}, {Brown}, {Chadwick}, {Chounet}, {Cornils},
  {Costamante}, {Degrange}, {Dickinson}, {Djannati-Ata{\"i}}, {O'C.~Drury},
  {Dubus}, {Emmanoulopoulos}, {Espigat}, {Feinstein}, {Fontaine}, {Fuchs},
  {Funk}, {Gallant}, {Giebels}, {Gillessen}, {Glicenstein}, {Goret},
  {Hadjichristidis}, {Hauser}, {Heinzelmann}, {Henri}, {Hermann}, {Hinton},
  {Hofmann}, {Holleran}, {Horns}, {Jacholkowska}, {de Jager}, {Kh{\'e}lifi},
  {Komin}, {Konopelko}, {Latham}, {Le Gallou}, {Lemi{\`e}re},
  {Lemoine-Goumard}, {Leroy}, {Lohse}, {Martin}, {Martineau-Huynh},
  {Marcowith}, {Masterson}, {McComb}, {de Naurois}, {Nolan}, {Noutsos},
  {Orford}, {Osborne}, {Ouchrif}, {Panter}, {Pelletier}, {Pita},
  {P{\"u}hlhofer}, {Punch}, {Raubenheimer}, {Raue}, {Raux}, {Rayner}, {Reimer},
  {Reimer}, {Ripken}, {Rob}, {Rolland}, {Rowell}, {Sahakian}, {Saug{\'e}},
  {Schlenker}, {Schlickeiser}, {Schuster}, {Schwanke}, {Siewert}, {Sol},
  {Spangler}, {Steenkamp}, {Stegmann}, {Tavernet}, {Terrier}, {Th{\'e}oret},
  {Tluczykont}, {Vasileiadis}, {Venter}, {Vincent}, {V{\"o}lk}, \&
  {Wagner}}]{ah05}
{Aharonian}, F., {Akhperjanian}, A.~G., {Bazer-Bachi}, A.~R., {et~al.} 2005,
  \aap, 437, L7

\bibitem[{{Aharonian} {et~al.}(2006){Aharonian}, {Akhperjanian}, {Bazer-Bachi},
  {Beilicke}, {Benbow}, {Berge}, {Bernl{\"o}hr}, {Boisson}, {Bolz}, {Borrel},
  {Braun}, {Breitling}, {Brown}, {Chadwick}, {Chounet}, {Cornils},
  {Costamante}, {Degrange}, {Dickinson}, {Djannati-Ata{\"i}}, {O'C.~Drury},
  {Dubus}, {Emmanoulopoulos}, {Espigat}, {Feinstein}, {Fontaine}, {Fuchs},
  {Funk}, {Gallant}, {Giebels}, {Glicenstein}, {Goret}, {Hadjichristidis},
  {Hauser}, {Hauser}, {Heinzelmann}, {Henri}, {Hermann}, {Hinton}, {Hofmann},
  {Holleran}, {Horns}, {Jacholkowska}, {de Jager}, {Kh{\'e}lifi}, {Klages},
  {Komin}, {Konopelko}, {Latham}, {Le Gallou}, {Lemi{\`e}re},
  {Lemoine-Goumard}, {Lohse}, {Martin}, {Martineau-Huynh}, {Marcowith},
  {Masterson}, {McComb}, {de Naurois}, {Nedbal}, {Nolan}, {Noutsos}, {Orford},
  {Osborne}, {Ouchrif}, {Panter}, {Pelletier}, {Pita}, {P{\"u}hlhofer},
  {Punch}, {Raubenheimer}, {Raue}, {Rayner}, {Reimer}, {Reimer}, {Ripken},
  {Rob}, {Rolland}, {Rowell}, {Sahakian}, {Saug{\'e}}, {Schlenker},
  {Schlickeiser}, {Schuster}, {Schwanke}, {Siewert}, {Sol}, {Spangler},
  {Steenkamp}, {Stegmann}, {Superina}, {Tavernet}, {Terrier}, {Th{\'e}oret},
  {Tluczykont}, {van Eldik}, {Vasileiadis}, {Venter}, {Vincent}, {V{\"o}lk}, \&
  {Wagner}}]{ah06}
{Aharonian}, F., {Akhperjanian}, A.~G., {Bazer-Bachi}, A.~R., {et~al.} 2006,
  \aap, 449, 223

\bibitem[{Berezhko {et~al.}(2002)Berezhko, Ksenofontov, \& V{\"{o}}lk}]{bk02}
Berezhko, E.~G., Ksenofontov, L.~T., \& V{\"{o}}lk, H.~J. 2002, A$\&$A, 395,
  943

\bibitem[{{Cargill} \& {Papadopoulos}(1988)}]{cp88}
{Cargill}, P.~J. \& {Papadopoulos}, K. 1988, \apjl, 329, L29

\bibitem[{{Carter} \& {Read}(2007)}]{cr07}
{Carter}, J.~A. \& {Read}, A.~M. 2007, \aap, 464, 1155

\bibitem[{Chevalier(1982)}]{ch82}
Chevalier, R.~A. 1982, ApJ, 258, 790

\bibitem[{Dubner {et~al.}(2002)Dubner, Giacani, Goss, Green, \& Nyman}]{dg02}
Dubner, G.~M., Giacani, E.~B., Goss, W.~M., Green, A.~J., \& Nyman, L.~A. 2002,
  A$\&$A, 387, 1047

\bibitem[{Dwarkadas \& Chevalier(1998)}]{dc98}
Dwarkadas, V.~V. \& Chevalier, R.~A. 1998, ApJ, 497, 807

\bibitem[{Dyer {et~al.}(2001)Dyer, Reynolds, Borkowski, Allen, \& Petre}]{dr01}
Dyer, K.~K., Reynolds, S.~P., Borkowski, K.~J., Allen, G.~E., \& Petre, R.
  2001, ApJ, 551, 439

\bibitem[{{Ghavamian} {et~al.}(2007){Ghavamian}, {Laming}, \&
  {Rakowski}}]{gl07}
{Ghavamian}, P., {Laming}, J.~M., \& {Rakowski}, C.~E. 2007, \apjl, 654, L69

\bibitem[{Ghavamian {et~al.}(2002)Ghavamian, Winkler, Raymond, \& Long}]{gw02}
Ghavamian, P., Winkler, P.~F., Raymond, J.~C., \& Long, K.~S. 2002, ApJ, 572,
  888

\bibitem[{{Hamilton} {et~al.}(1986){Hamilton}, {Sarazin}, \&
  {Szymkowiak}}]{hs86}
{Hamilton}, A.~J.~S., {Sarazin}, C.~L., \& {Szymkowiak}, A.~E. 1986, \apj, 300,
  698

\bibitem[{{Heng} \& {McCray}(2007)}]{hm07}
{Heng}, K. \& {McCray}, R. 2007, \apj, 654, 923

\bibitem[{{Iwamoto} {et~al.}(1999){Iwamoto}, {Brachwitz}, {Nomoto},
  {Kishimoto}, {Umeda}, {Hix}, \& {Thielemann}}]{ib99}
{Iwamoto}, K., {Brachwitz}, F., {Nomoto}, K., {et~al.} 1999, \apjs, 125, 439

\bibitem[{{Korreck} {et~al.}(2004){Korreck}, {Raymond}, {Zurbuchen}, \&
  {Ghavamian}}]{kr04}
{Korreck}, K.~E., {Raymond}, J.~C., {Zurbuchen}, T.~H., \& {Ghavamian}, P.
  2004, \apj, 615, 280

\bibitem[{Koyama {et~al.}(1995)Koyama, Petre, Gotthelf, Hwang, Matsuura, Ozaki,
  \& Holt}]{ko95}
Koyama, K., Petre, R., Gotthelf, E.~V., {et~al.} 1995, Nature, 378, 255

\bibitem[{{Ksenofontov} {et~al.}(2005){Ksenofontov}, {Berezhko}, \&
  {V{\"o}lk}}]{kb05}
{Ksenofontov}, L.~T., {Berezhko}, E.~G., \& {V{\"o}lk}, H.~J. 2005, \aap, 443,
  973

\bibitem[{{Laming} {et~al.}(1996){Laming}, {Raymond}, {McLaughlin}, \&
  {Blair}}]{lr96}
{Laming}, J.~M., {Raymond}, J.~C., {McLaughlin}, B.~M., \& {Blair}, W.~P. 1996,
  \apj, 472, 267

\bibitem[{{Lodders}(2003)}]{lo03}
{Lodders}, K. 2003, \apj, 591, 1220

\bibitem[{{Long} {et~al.}(2003){Long}, {Reynolds}, {Raymond}, {Winkler},
  {Dyer}, \& {Petre}}]{lr03}
{Long}, K.~S., {Reynolds}, S.~P., {Raymond}, J.~C., {et~al.} 2003, \apj, 586,
  1162

\bibitem[{{Moffett} {et~al.}(1993){Moffett}, {Goss}, \& {Reynolds}}]{mg93}
{Moffett}, D.~A., {Goss}, W.~M., \& {Reynolds}, S.~P. 1993, \aj, 106, 1566

\bibitem[{{Parizot} {et~al.}(2006){Parizot}, {Marcowith}, {Ballet}, \&
  {Gallant}}]{pm06}
{Parizot}, E., {Marcowith}, A., {Ballet}, J., \& {Gallant}, Y.~A. 2006, \aap,
  453, 387

\bibitem[{{Raymond} {et~al.}(2007){Raymond}, {Korreck}, {Sedlacek}, {Blair},
  {Ghavamian}, \& {Sankrit}}]{rk07}
{Raymond}, J.~C., {Korreck}, K.~E., {Sedlacek}, Q.~C., {et~al.} 2007, \apj,
  659, 1257

\bibitem[{Reynolds \& Gilmore(1986)}]{rg86}
Reynolds, S.~P. \& Gilmore, D.~M. 1986, AJ, 92, 1138

\bibitem[{Rothenflug {et~al.}(2004)Rothenflug, Ballet, Dubner, Giacani,
  Decourchelle, \& Ferrando}]{rb04}
Rothenflug, R., Ballet, J., Dubner, G., {et~al.} 2004, A$\&$A, 425, 121

\bibitem[{Sgro(1975)}]{sg75}
Sgro, A. 1975, ApJ, 197, 621

\bibitem[{{Truelove} \& {McKee}(1999)}]{tm99}
{Truelove}, J.~K. \& {McKee}, C.~F. 1999, \apjs, 120, 299

\bibitem[{Wang \& Chevalier(2001)}]{wc01}
Wang, C.~Y. \& Chevalier, R.~A. 2001, ApJ, 549, 1119

\bibitem[{Winkler {et~al.}(2003)Winkler, Gupta, \& Long}]{wg03}
Winkler, P.~F., Gupta, G., \& Long, K.~S. 2003, ApJ, 585, 324

\bibitem[{Winkler \& Long(1997)}]{wl97}
Winkler, P.~F. \& Long, K.~S. 1997, ApJ, 491, 829

\end{thebibliography}
\end{document}